\documentclass{article}
\usepackage{ltwol2e}

\usepackage{epsfig}

\def\Journal#1#2#3#4{{#1} {\bf #2}, #3 (#4)}

\def\MPA{{\em Mod. Phys. Lett.} A}
\def\NPB{{\em Nucl. Phys.} B}
\def\PLB{{\em Phys. Lett.}  B}

\def\PRD{{\em Phys. Rev.} D}

\def\ffbar{\relax\ifmmode{\mathrm{f}\overline{\mathrm{f}}}%
          \else${\mathrm{f}\overline{\mathrm{f}}}$\fi}
\def\qqbar{\relax\ifmmode{\mathrm{q}\overline{\mathrm{q}}}%
          \else${\mathrm{q}\overline{\mathrm{q}}}$\fi}
\def\llbar{\relax\ifmmode{\ell^+\ell^-}
          \else${\ell^+\ell^-}$\fi}
\def\ttbar{\relax\ifmmode{\tau^+\tau^-}
          \else${\tau^+\tau^-}$\fi}
\def\ee{\relax\ifmmode{{\mathrm e}^+{\mathrm e}^-}
          \else{e$^+$e$^-$}\fi}
\def\nunu{\relax\ifmmode{\nu\nu}
          \else{$\nu\nu$}\fi}
\def\mm{\relax\ifmmode{\mu^+\mu^-}
          \else{$\mu^+\mu^-$}\fi}
\def\tt{\relax\ifmmode{\tau^+\tau^-}
          \else{$\tau^+\tau^-$}\fi}
\def\WW{\relax\ifmmode{{\mathrm W}^+{\mathrm W}^-}
          \else{W$^+$W$^-$}\fi}
\def\ppbar{\relax\ifmmode{\mathrm{p}\overline{\mathrm{p}}}%
          \else${\mathrm{p}\overline{\mathrm{p}}}$\fi}
\def\bbbar{\relax\ifmmode{\mathrm{b}\overline{\mathrm{b}}}%
          \else${\mathrm{b}\overline{\mathrm{b}}}$\fi}
\def\ccbar{\relax\ifmmode{\mathrm{c}\overline{\mathrm{c}}}%
          \else${\mathrm{c}\overline{\mathrm{c}}}$\fi}
\def\GZ{\relax\ifmmode{\Gamma_{\PZ}}%
          \else$\Gamma_{\PZ}$\fi}
\def\MZ{\relax\ifmmode{m_{\PZ}}%
          \else$m_{\PZ}$\fi}

\newcommand{\shat}{\hat{s}}

\def\bea{\begin{eqnarray}}
\def\eea{\end{eqnarray}}

\begin{document}

\begin{titlepage}
  \begin{center}
    \font\GIANT=cmr17 scaled\magstep4
       {\GIANT U\kern0.8mm N\kern0.8mm I\kern0.8mm V\kern0.8mm %
        E\kern0.8mm R\kern0.8mm S\kern0.8mm I\kern0.8mm %
        T\kern0.8mm %
         \setbox0=\hbox{A}\setbox1=\hbox{.}%
         \dimen0=\ht0 \advance \dimen0 by -\ht1%
         \makebox[0mm][l]%
         {\raisebox{\dimen0}{.\kern 0.2\wd0 .}}A\kern0.8mm
            T\kern5mm
            B\kern0.8mm O\kern0.8mm N\kern0.8mm N\kern0.8mm
         } \\[8mm]
       {\GIANT
        P\kern0.8mm h\kern0.8mm y\kern0.8mm s\kern0.8mm i\kern0.8mm
        k\kern0.8mm a\kern0.8mm l\kern0.8mm i\kern0.8mm s\kern0.8mm
        c\kern0.8mm h\kern0.8mm e\kern0.8mm s\kern5mm
        I\kern0.8mm n\kern0.8mm s\kern0.8mm t\kern0.8mm
        i\kern0.8mm t\kern0.8mm u\kern0.8mm t\kern0.8mm
       } \\[2.5cm]
       {\Large \bf
Single and Pair Production of Neutral Electroweak 
Gauge Bosons at LEP}
\\ [15mm]

    {\Large
      Michael Kobel\\}
{\large \it
Physikalisches Institut, Universit{\"a}t Bonn, D-53115 Bonn, FRGermany\\
on leave of absence from Fakult{\"a}t f{\"u}r Physik, Universit{\"a}t Freiburg,
D-79104 Freiburg, FRGermany \\E-mail: Michael.Kobel@cern.ch\\
representing the LEP collaborations at ICHEP-98, Vancouver.}   
\\[20mm]  
  \end{center}

\noindent
\large
  Recent LEP results on single and pair production of neutral electroweak
gauge bosons are reviewed.  QED and Electroweak $\gamma$-e Compton scattering
at LEP covers $\gamma$-e center-of-mass energies
$\sqrt{\shat}$ in the range from about 20~GeV to 170~GeV,
and leads to single production of on-shell $\gamma$,
off-shell $\gamma^*$, and Z bosons, also known as ``Zee'' process.
The latter two final states have 
been observed for the first time by the OPAL collaboration, 
while the measurement of the scattered on-shell $\gamma$'s by L3
represents the highest energies at which QED Compton scattering
has been studied so far. These processes can be used to set limits on excited
electrons. Pair production of $\gamma^*$ and/or Z 
at the \ee\ center-of-mass energy $\sqrt{s}$=183~GeV has
been studied by the DELPHI, L3, and OPAL collaborations. The
combination of these experiments yields the first significant
measurement of Z pair production. With more statistics at higher energies, 
interesting limits on anomalous $\gamma$ZZ and ZZZ couplings can be
derived from this process.

\noindent
%
  \begin{figure}[b]
    \begin{minipage}{\textwidth}
        \begin{raggedright}
        \begin{tabular}{@{}l@{}}
                Post address:  \\
                Nussallee 12   \\
                D-53115 Bonn   \\
                Germany      \\
        \end{tabular}
        \end{raggedright}
        \hfill
        \parbox{5.5cm}
        {
            \vbox to 2cm
            {
                \vskip -1.4truecm
                \hbox to 5.5cm
                {
                     \includegraphics{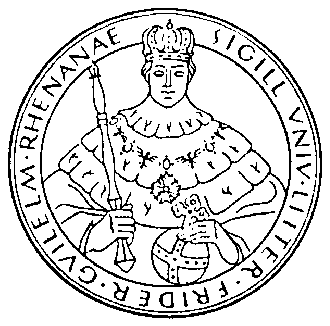}
                    \hfill
                }
                \vfill
                \centering
            }
        }
        \hfill
        \begin{raggedleft}
        \begin{tabular}{@{}l@{}}
            BONN-HE-98-05                   \\          
            Bonn University                 \\
            October 1998                       \\          
        \end{tabular}
        \end{raggedleft}
    \end{minipage}
  \end{figure}
\end{titlepage}
\newpage
\title{SINGLE AND PAIR PRODUCTION OF NEUTRAL ELECTROWEAK 
GAUGE BOSONS AT LEP}

\author{Michael Kobel}

\address{Physikalisches Institut, Universit{\"a}t Bonn, D-53115 Bonn, FRGermany\\
on leave of absence from Fakult{\"a}t f{\"u}r Physik, Universit{\"a}t Freiburg,
79104 Freiburg, FRGermany \\E-mail: Michael.Kobel@cern.ch}   

\author{representing the LEP collaborations}

\twocolumn[\maketitle\abstracts{
Recent LEP results on single and pair production of neutral electroweak
gauge bosons are reviewed.  QED and Electroweak $\gamma$-e Compton scattering
at LEP covers $\gamma$-e center-of-mass energies
$\sqrt{\shat}$ in the range from about 20~GeV to 170~GeV,
and leads to single production of on-shell $\gamma$,
off-shell $\gamma^*$, and Z bosons, also known as ``Zee'' process.
The latter two final states have 
been observed for the first time by the OPAL collaboration, 
while the measurement of the scattered on-shell $\gamma$'s by L3
represents the highest energies at which QED Compton scattering
has been studied so far. These processes can be used to set limits on excited
electrons. Pair production of $\gamma^*$ and/or Z 
at the \ee\ center-of-mass energy $\sqrt{s}$=183~GeV has
been studied by the DELPHI, L3, and OPAL collaborations. The
combination of these experiments yields the first significant
measurement of Z pair production. With more statistics at higher energies, 
interesting limits on anomalous $\gamma$ZZ and ZZZ couplings can be
derived from this process.   
}]

\section{Single Gauge Boson Production in quasi-real Compton Scattering}

\begin{figure}
\center
\epsfig{file=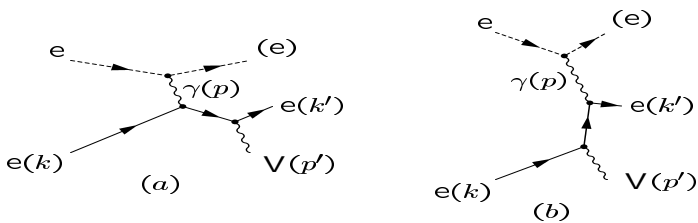,width=0.5\textwidth,height=2.8cm}
\caption{Diagrams for the process ee$\to$(e)eV.}
\label{fig:zee}
\end{figure}

The elementary subprocess for the production of a
single neutral gauge boson V = ($\gamma$, $\gamma^*$, Z)  
via Compton Scattering at LEP is e$\gamma\to$eV, 
where a quasi-real photon $\gamma$, radiated from one
of the beam electrons, scatters off the other electron e, as shown
in Fig.~\ref{fig:zee}. The electron (e) that radiated the
incoming quasi-real photon usually stays unobserved close to the beamline.

\noindent
For real incoming photons ($p^2=0$), 
the cross-section dependence of the process
e$(k)\gamma(p) \to$ e$(k^\prime)$V$(p\prime)$
on the Mandelstam variables
$\hat{s}=(k^\prime+p^\prime)^2$, $\hat{t}=(k^\prime-k)^2$,
$\hat{u}=(p^\prime-k)^2$ is~\cite{EPSINGLEZ} 
\begin{equation}
\frac{{\rm d}\sigma}{{\rm d}\hat{t}} \propto
\frac{1}{\hat{s}^2}\left(
        \frac{\hat{u}}{\hat{s}} +
        \frac{2{p^\prime}^2\hat{t}}{\hat{u}\hat{s}} +
        \frac{\hat{s}}{\hat{u}} \right).
\label{eq:compton}
\end{equation} 
For ${p^\prime}^2 = 0$ the well-known terms for QED Compton
scattering remain. 

\subsection{QED Compton Scattering, V=$\gamma$}\label{subsec:l3c}
\begin{figure}
\center
\epsfig{file=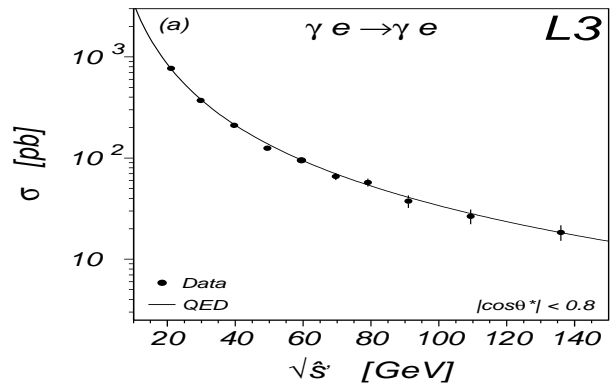,width=0.45\textwidth,height=5.cm}
\caption{Measured total cross-section of QED Compton Scattering
as a function of  $\sqrt{\shat}$. The solid line shows the QED prediction.}
\label{fig:L3_shat}
\end{figure}

\noindent
The L3 collaboration has observed 4641 candidate events for
the e$\gamma$ final state produced in QED Compton Scattering~\cite{L3comp},
collected in 230~pb$^{-1}$ of data taken at \ee\ center-of-mass energies
between 91 and 183~GeV. 
Background from \ee$\to$\ee\ and \ee$\to\gamma\gamma$ has been reduced to 
below 0.5\% by requiring $E_2$, the energy of the
lower energetic cluster of the e$\gamma$ pair to be less than 85\% of the
beam energy. Within $|\cos\theta^*|<0.8$ the cross-section as a function
of the e$\gamma$ centre-of mass energy  $\sqrt{\shat}$ 
has been measured in the range from 20~GeV to 140~GeV, 
which is the highest energy at which QED Compton Scattering  
has been studied so far. The measured cross-section 
is found to be in agreement with the QED prediction
(see Fig.~\ref{fig:L3_shat}).

\begin{figure*}[htb]
\begin{center}
\epsfig{file=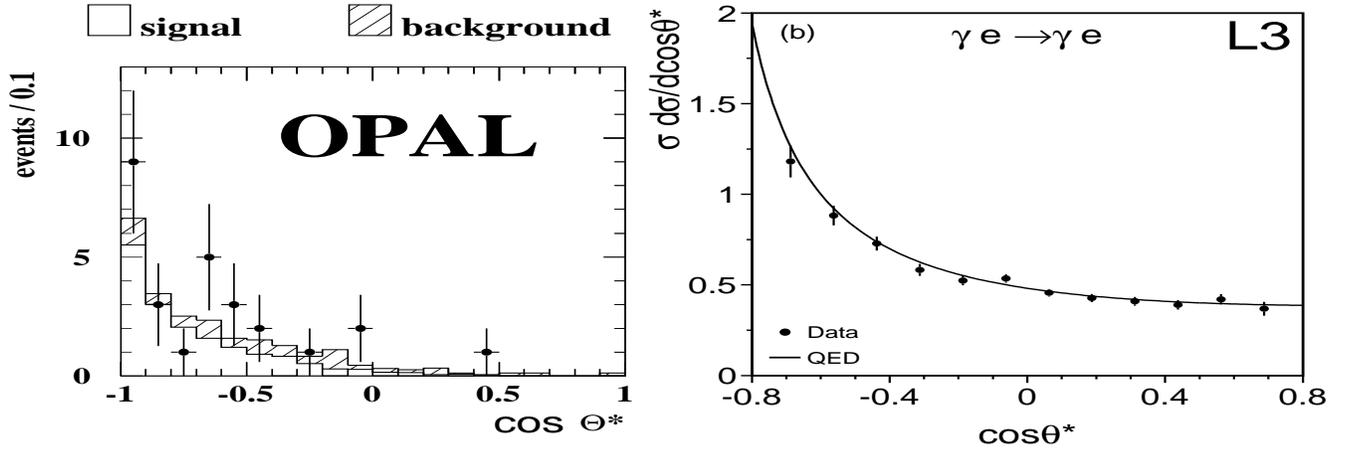,width=1.0\textwidth,height=6cm}
\end{center}
\caption{Distributions of $\cos\theta^*$ for QED Compton scattering (right)
and Electroweak Compton Scattering (left).
}
\label{fig:costhstar}
\end{figure*}

\subsection{Elektroweak Compton Scattering, V=$\gamma*$,Z}\label{subsec:opalc}

\begin{figure}
\center
\epsfig{file=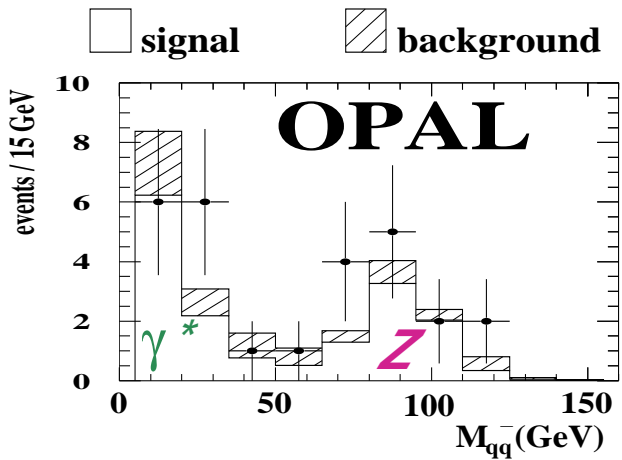,width=0.45\textwidth,height=6.cm}
\caption{Invariant mass distribution of the hadronic system
in candidate events for the process \ee$\to$(e)e$\gamma^*$ and (e)eZ.
}
\label{fig:opal_mqq}
\end{figure}

\noindent
The production of Z bosons and virtual $\gamma^*$'s with a mass 
$\sqrt{{p^\prime}^2}>$ 5~GeV 
decaying to hadronic final states has been studied
by the OPAL collaboration~\cite{OPALcomp} 
using 55~pb$^{-1}$ of data taken at an 
\ee\ center-of-mass energy of 183~GeV.
The final state is characterized by a single, usually low-energy,
electron, isolated from a hadronic system containing one or two jets
from the Z/$\gamma^*$ decay. The large background of multihadronic 
\ee$\to$\qqbar, semileptonic WW$\to$qqe$\nu$, and especially of events
from high q$^2$ e-$\gamma$ deep inelastic scattering is greatly reduced 
by fake electron rejection and by
exploiting the distinct signal kinematics, i.e. the scattering angles
of the final state particles e and V, and the missing momentum 
due to the escaping electron (e) along the beam. 

After all cuts, 27 candidate events remain in the data, 14 in a hadronic
mass range of 5~GeV$ < m_\qqbar <$60~GeV, dominated by the (e)e$\gamma^*$
final state, and 13 above 60~GeV, dominated by (e)eZ. The expected
backgrounds are 4.4 and 2.1 events, respectively. The observed
significant excess in both mass ranges constitutes the first observation 
of the processes \ee$\to$(e)e$\gamma^*$ and (e)eZ.
The respective contributions of $\gamma^*$ and Z 
are clearly visible in the hadronic mass
distribution, shown in Fig.~\ref{fig:opal_mqq}.
The measured cross-section within the detector acceptance is
($4.1 \pm 1.6 \pm 0.6$)pb in the ``(e)e$\gamma^*$'' region and 
($0.9 \pm 0.3 \pm 0.1$)pb in the ``(e)eZ'' region, in good agreement
with the predictions of two Monte Carlo generators for this process, 
PYTHIA and grc4f.

\subsection{Limits on excited Electrons}
\noindent
The typical transverse momentum of the
scattered boson V is small ($\hat{u} \to 0$ in Eq.~\ref{eq:compton}),
leading to a dominance of the diagram Fig.~\ref{fig:zee}b.  Its kinematic
is characterized by a hard spectrum of the boson V, 
emitted preferentially close to the initial direction $\cos\theta^*=-1$
of the scattered electron in the e-$\gamma$ rest system, as 
shown for both, QED and Elektroweak Compton scattering,
in  Fig.~\ref{fig:costhstar}.

\begin{figure*}
\vspace*{-3.5cm}   
\begin{center}
\epsfig{file=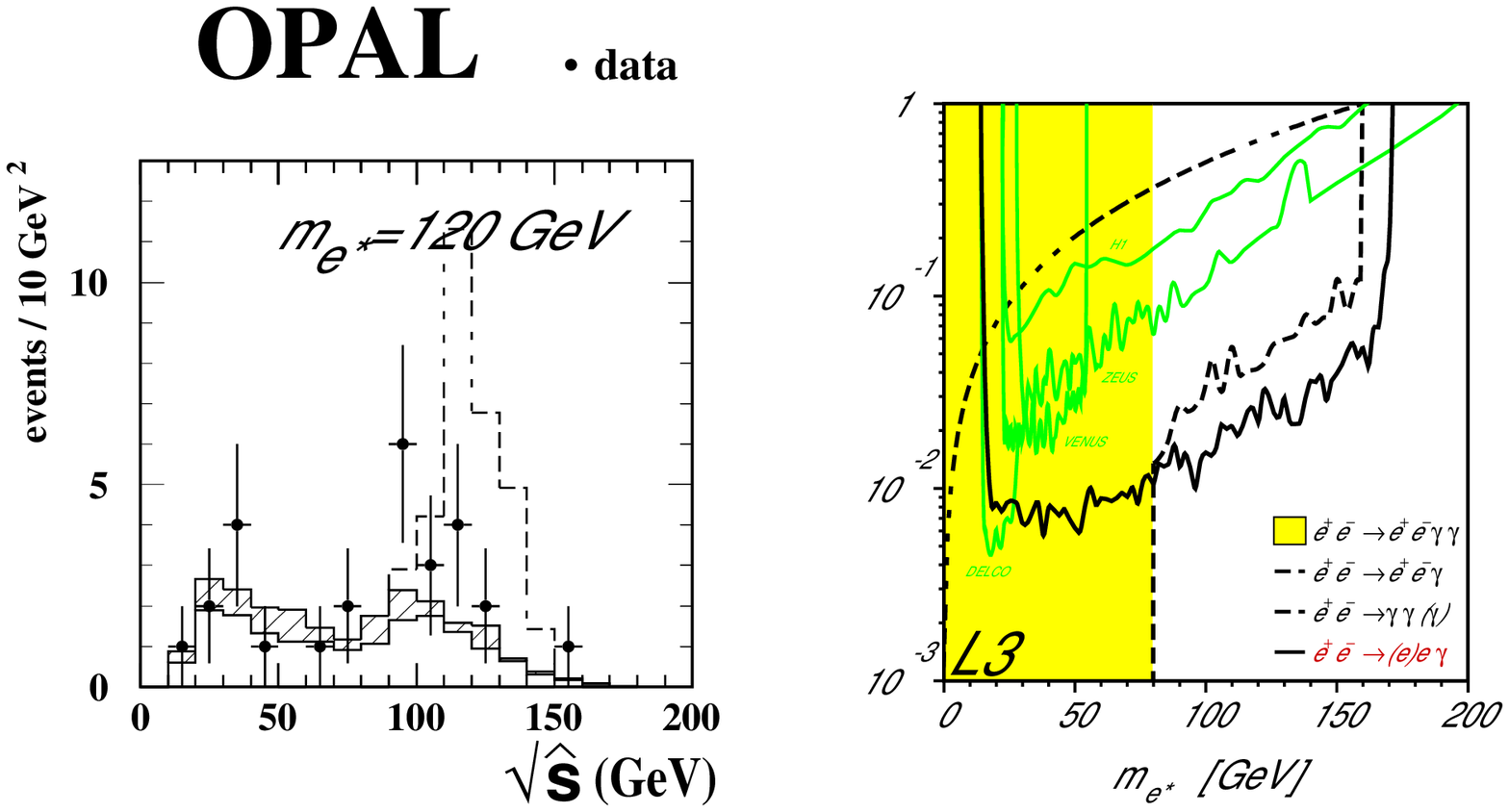,width=0.95\textwidth,height=11cm}
\end{center}
\vspace*{-1.5cm}   
\caption{
Left-hand side: Distribution of the e-$\gamma$ centre-of-mass 
energy in candidate events for Electroweak Compton scattering.
The dashed histogram shows a hypothetical e$^*\to$eZ signal. 
Right-hand side: Limits in units of $e$ for  e$^*$-e$\gamma$ coupling
as a function of the  e$^*$ mass, obtained from 
QED Compton Scattering. 
}
\label{fig:estar}
\end{figure*}

The diagram Fig.~\ref{fig:zee}a could be
enhanced by contributions of a hypothetical excited electron e$^*$,
coupling to e$\gamma$ and/or to eZ. This would manifest itself 
as a peak in the $\sqrt{\shat}$ distribution at the e$^*$ mass.
From the absence of such a peak in Fig.~\ref{fig:L3_shat}
the L3 collaboration obtains limits on the free coupling parameter
$\lambda$ in the e$^*$-e$\gamma$ coupling, in units of the
electron charge, $e$, shown as the solid line on the right-hand plot of  
Fig.~\ref{fig:estar}.
They improve existing limits above the exclusion limit from e$^*$ pair
production.

A hypothetical signal for  e$^*\to$eZ in Electroweak Compton Scattering
with a production cross-section times
branching ratio of 5~pb is shown as dashed line 
in the left-hand plot of Fig.~\ref{fig:estar}.
Quantitative limits for excited electrons
decaying to eZ,$\nu$W, and e$\gamma$ are given in
another contribution to the conference~\cite{teuscher}.

\section{Pair Production of $\gamma*$ and/or Z bosons}

Neutral current 4-fermion production, where all fermions are visible under
large angles to the beamline, is dominated by the so-called ``conversion''
diagram shown in Fig.\ref{fig:zz}. The data taken at the kinematic threshold of
$\sqrt{s}=183$~GeV offer the first opportunity to  detect  on-shell 
Z boson pair production via this diagram,  by applying mass constraints
on the fermion pairs. Some final states of
Z pair production are very hard to
separate from the production of a Standard Model Higgs Boson close to the 
Z mass. In case of an excess, only  a coverage of a large variety of 
decay channels will help to disentangle Higgs production from possible
anomalous $\gamma$ZZ and ZZZ couplings.
Inclusive  neutral current
4-fermion production without any  mass constraints,   
mainly proceeding via the $\gamma^*$Z intermediate state, has  been studied
in addition. It's observation is an important test for 
the understanding of higher order processes in the Standard Model, which 
often constitute a prominent background in  the search for new physics.

\begin{figure}
\center
\caption{Pair Production of $\gamma*$ and/or Z bosons.}
\label{fig:zz}
\end{figure}

\subsection{Inclusive Neutral Current 4-Fermions}

The DELPHI~\cite{delphi-4f} and L3~\cite{l3-4f} collaborations
have updated their inclusive studies of Neutral Current 4-fermion events in the
final states $\ell\ell$\qqbar\ and $\ell\ell\ell\ell$, as detailed in 
Table~\ref{tab:4f-incl}. There is a tendency of observing rather 
more  $\ell\ell$qq events than predicted. 
Since systematic errors have not yet been determined,
no statement about the significance of this trend can presently be made.

\begin{table}
\begin{center}
\caption{ Preliminary values for number of observed events, expected signal 
and background for  inclusive neutral current 4-fermion processes
at $\sqrt{s}$=183~GeV. Only statistical errors are given.}\label{tab:4f-incl}
\vspace{0.2cm}
\begin{tabular}{|c|c|c|c|}
\hline
Channel(Exp.) & $N_{\rm obs}$ & $N_{\rm sig}$ &  $N_{\rm bck}$ \\
\hline
eeqq (DELPHI)           & 3  & 2.9$\pm$0.2 & 0.6$\pm$0.2 \\ 
$\mu\mu$qq (DELPHI)     & 8  & 2.9$\pm$0.2 & 0.4$\pm$0.2 \\ 
$\ell\ell$qq (L3)       & 13 & 7.4$\pm$0.4 & 2.0$\pm$0.1 \\ 
\hline
$\ell\ell\ell\ell$ (DELPHI) & 3  & 3.7$\pm$0.2 & 0.1$\pm$0.2 \\ 
$\ell\ell\ell\ell$ (L3)     & 11 & 4.4$\pm$0.1 & 4.2$\pm$0.9 \\
\hline 
\end{tabular}
\end{center}
\end{table}

\subsection{Selection of Z pair events}

Z pair production has been searched for~\cite{delphi-4f,l3-4f,opalzz} 
at $\sqrt{s}$ = 183~GeV in the final 
states (Branching ratio, experiments)
qqqq (49\%, L3 and OPAL), qqbb (19\%, DELPHI and OPAL),
qq$\nu\nu$ (28\%, L3 and OPAL), qqee (5\%, DELPHI. L3 and OPAL),
qq$\mu\mu$ (5\%, DELPHI, L3 and OPAL), qq$\tau\tau$ (5\%, OPAL),
and $\ell\ell\ell\ell$ (1\%, DELPHI and OPAL).
A significant measurement is very difficult, since the expected total 
cross-section of 0.24~pb for the pure ZZ diagram
is more than 60 times lower
than that of W pair production,  the most prominent background
for the final states with large branching ratio. On the other hand, the
cleaner channels like qqee and qq$\mu\mu$ have small branching ratios,
so that for each such channel
only about 1 event is expected per experiment with the
available integrated luminosity of about 55~pb$^{-1}$. 

\begin{figure}
\center
\epsfig{file=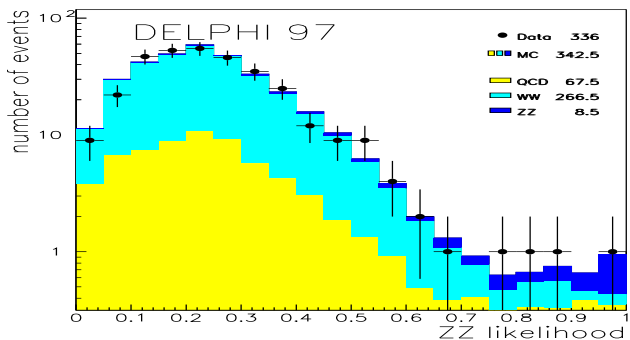,height=5cm,width=0.5\textwidth}
\caption{Likelihood distribution for the ZZ$\to$qqbb final state.}
\label{fig:qqbb_delphi}
\end{figure}

The hadronic channels qqqq and qq$\nu\nu$ have been analysed using
likelihood techniques and neural nets. DELPHI and OPAL have in addition
studied the subsample of qqbb, where the b-tag helps in reducing
the overwhelming W pair background. An example of a likelihood distribution
for the qqbb channel from the DELPHI experiment is shown in 
Fig.~\ref{fig:qqbb_delphi}

In the $\nu\nu$qq final state
there is important additional background 
from single W production, (e)$\nu$W, and double initial
state radiation return to the Z, ($\gamma\gamma$)Z, both having 
even after preselection cuts, cross-sections comparable to the signal.
All experiments observe more events in the data than expected 
from pure background, but --- with exception from DELPHI,
who apparently profit from a statistical upward fluctuation  ---
the excess is not large enough to significantly claim to have observed
Z-pair production from one experiment alone.

\subsection{LEP average of Z pair cross-section}

\begin{figure}
\center
\epsfig{file=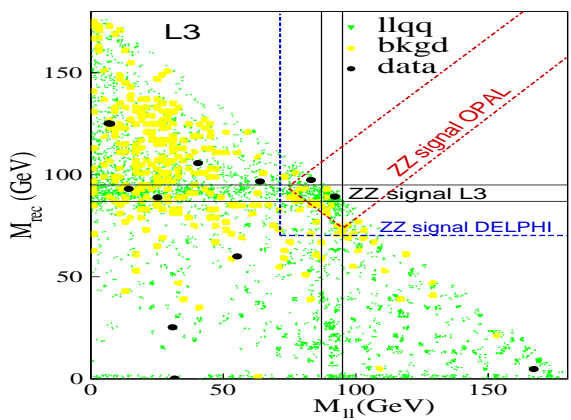,width=0.45\textwidth,height=6cm}
\caption{Pair Production of $\gamma*$ and/or Z bosons.}
\label{fig:zzdef}
\end{figure}

The average of the observations from the LEP experiments
is not straight forward, since very different signal definitions
were adopted in terms of the invariant masses of the two
fermion pairs.
They are indicated in Fig.\ref{fig:zzdef}, which shows
the L3 result for the combined eeqq+$\mu\mu$qq final state.

For performing the average I adopted for each channel, $i$, the cross-section of
the pure ZZ diagram, $\sigma_i$(ZZ), integrated
over the whole phase space, as the signal cross-section to be measured.
The theoretical prediction for the sum of all channels,
$\sigma$(ZZ)=0.24~pb at an average $\sqrt{s}=182.7$~GeV,
was determined using the YFSZZ generator~\cite{YFSZZ}.
The number of observed events in the signal region, as defined by each
experiment, the selection efficiency~\footnote{
DELPHI and L3 have given no numbers for selection efficiencies.
They instead give the number of expected signal and background events
with statistical errors, only. Their signal definition comprises
all neutral current 4-fermion diagrams for each final state
within their respective fermion pair mass regions. 
I estimated their  efficiency by
determining the corresponding expected cross-sections with the help 
of the appropriate Monte Carlo Generators.
This procedure resulted in relative efficiency errors of 
${\cal O}$(10--15\%), assumed to be large enough to
cover the missing systematic errors.
}
 within this signal region,
and the expected background events were then subjected to a
maximum likelihood fit with $\sigma$(ZZ) as the only free parameter.
The qqqq and qq$\nunu$ channels of the L3 experiment were excluded
from the average, since there  
the signal had not been defined in terms of the pair masses. 
Due to large background in these channels their contribution
to the average would have been in any case very limited.
To perform the likelihood fit, the branching ratio of each channel $i$
in each experiment $j$ was scaled by a factor
$$ f_{ij} = \frac{\sigma_{ij}(ZZ+Z\gamma^*+\gamma^*Z+eeZ)}{\sigma_i(ZZ)} $$
where $\sigma_{ij}$ is the expected cross-section for all diagrams
leading to the final state $i$, (dominated by those listed in parentheses),
for the signal definition of experiment $j$.
The scaling factors obtained from MC predictions
for the respective processes are listed in Table~\ref{tab:fscale}.
Values much lower than unity indicate a restrictive signal
definition, while loose signal definitions lead to values larger
than unity,  especially for eeff final states, where a considerable
contribution from the eeZ diagrams of Fig.~\ref{fig:zee} can enter.

 \begin{table}
\begin{center}
\caption{Scaling factors for single channel branching ratios, 
as explained in the text.}
\label{tab:fscale}
\vspace{0.2cm}
\begin{tabular}{|c|l|l|l|}
\hline
channel & DELPHI & L3 & OPAL \\
\hline
$\ell\ell\ell\ell$      & 1.40 & 0.55 & 0.98 \\
eeqq                    & 1.42 & 0.54 & 1.01 \\ 
$\mu\mu$qq              & 1.05 & 0.52 & 0.89 \\
qqqq                    & 0.95 & 0.51 & 0.86 \\
\hline 
\end{tabular}
\end{center}
\end{table}

\begin{figure}
\center
\epsfig{file=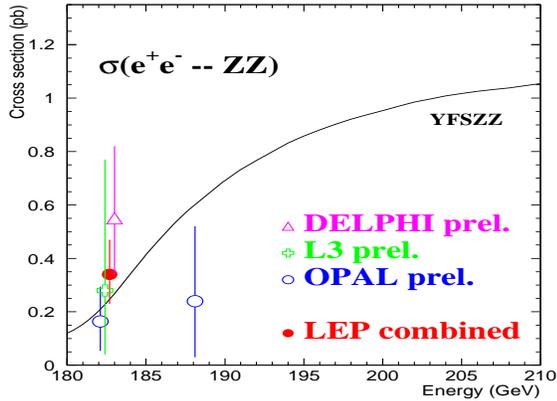,width=0.45\textwidth,height=6cm}
\caption{The results of the likelihood fit of ZZ cross-sections.
For better visibility a finite distribution of the points along
the energy axis is made.}
\label{fig:lepsec}
\end{figure}

Summed over the three experiments, a total number of 16 candidate
events, for about 9 expected background events, went in the fit.
The fit results are 
$\sigma(ZZ)$ = $0.63^{+0.31+0.06}_{-0.24-0.02}$~pb for
DELPHI, using 4 decay channels, 
$\sigma(ZZ)$ = $0.28^{+0.48+0.10}_{-0.24-0.02}$~pb for
L3, using 3 decay channels, and 
$\sigma(ZZ)$ = $0.16^{+0.17+0.03}_{-0.12-0.02}$~pb for
OPAL, using 7 decay channels, plus an extra
overlap channel between qqbb and qqqq.
The LEP average yields 
$$\sigma(ZZ) = 0.36^{+0.14+0.03}_{-0.12-0.02}{\rm pb},$$
constituting the first significant observation of ZZ production, 
at a rate consistent with expectation.
The results are 
plotted in Fig.~\ref{fig:lepsec}, where also a preliminary OPAL
measurement with the first 40~pb$^{-1}$ of data taken
at $\sqrt{s}$=189~GeV is indicated. These very recent data have also been
studied by DELPHI~\cite{delphi-189}, leading results consistent
with the Standard Model expectation.

\subsection{Anomalous VZZ Couplings (outlook)}

\begin{figure}
\center
\epsfig{file=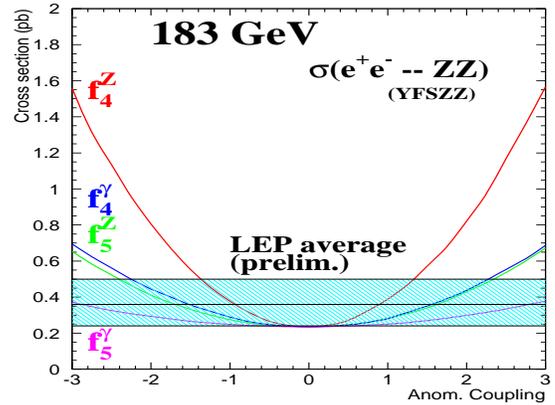,width=0.45\textwidth,height=6cm}
\caption{
Dependence of the predicted ZZ cross-section 
as a function of anomalous couplings,
and the measured preliminary LEP value. 
}
\label{fig:lepano}
\end{figure}

For on-shell ZZ production, there are 2 possible anomalous couplings,
called $f_4$ (CP-odd) and $f_5$ (CP-even)
for each, the $\gamma^*$ZZ and the Z$^*$ZZ vertex~\cite{hagiwara}.
They are zero in tree level Standard Model. Limiting these
with more data from LEP2 will pose nontrivial constraints especially on 
CP violating triple gauge couplings, like non-zero Z$^*$ZZ
vertices via Higgs loop effects~\cite{changkeungpal}.
From Fig.~\ref{fig:lepano} one obtains an impression of the current
sensitivity, assuming that the selection efficiency does
not depend on the anomalous coupling.

\noindent

\section*{Acknowledgements}
I would like to thank the organizers of teh conference
for their hospitality,
the representatives of the LEP experiments for 
their cooperation,
and especially David Strom for supplying me with the fitting software.

\end{document}